\documentclass[graybox]{svmult}

\usepackage{mathptmx}       
\usepackage{helvet}         
\usepackage{courier}        
\usepackage{type1cm}        
\usepackage{makeidx}         
\usepackage{graphicx}        
\usepackage{multicol}        
\usepackage[bottom]{footmisc}

\makeindex             

\def\spose#1{\hbox to 0pt{#1\hss}}
\def\gta{\mathrel{\spose{\lower 3pt\hbox{$\mathchar"218$}}
     \raise 2.0pt\hbox{$\mathchar"13E$}}}
\def\dt{\spose{\raise 1.0ex\hbox{\hskip2pt$\mathchar"201$}}}    
\newcommand{\msun}{\mbox{${\rm M}_\odot$}}
\newcommand{\mup}{m_{\rm up}}
\newcommand{\mlo}{m_{\rm lo}}
\newcommand{\Mup}{M_{\rm up}}
\newcommand{\Mlo}{M_{\rm lo}}

\begin{document}

\title*{The Salpeter Slope of the IMF Explained}
\author{M. S. Oey}
\institute{M. S. Oey \at Astronomy Department, University of Michigan,
830 Dennison Building, Ann Arbor, MI   48109-1042, \email{msoey@umich.edu}
}
%
%
\maketitle

\abstract*{
If we accept a paradigm that star formation is a self-similar,
hierarchical process, then the Salpeter slope of the IMF for high-mass
stars can be simply and elegantly explained as follows.  If the
instrinsic IMF at the smallest scales follows a simple --2 power-law
slope, then the steepening to the --2.35 Salpeter value results when
the most massive stars cannot form in the lowest-mass clumps of a
cluster.  It is stressed that this steepening {\bf must} occur if
clusters form hierarchically from clumps, and the
lowest-mass clumps can form stars.  This model is consistent with a
variety of observations as well as theoretical simulations. 
}

\abstract{
If we accept a paradigm that star formation is a self-similar,
hierarchical process, then the Salpeter slope of the IMF for high-mass
stars can be simply and elegantly explained as follows.  If the
instrinsic IMF at the smallest scales follows a simple --2 power-law
slope, then the steepening to the --2.35 Salpeter value results when
the most massive stars cannot form in the lowest-mass clumps of a
cluster.  It is stressed that this steepening {\bf must} occur if
clusters form hierarchically from clumps, and the
lowest-mass clumps can form stars.  This model is consistent with a
variety of observations as well as theoretical simulations. 
}

\section{Self-similar Hierarchical Fragmentation}


It is well known that at stellar masses $m \gta 1\ \msun$,
the initial mass function follows the Salpeter (1955) power-law slope:
\begin{equation}
N(m)\ dm \propto m^{-2.35}\ dm \quad .
\end{equation}
This represents the distribution of stellar birth masses, showing a
power-law index $\alpha = -2.35$, which is observed in most
massive star-forming environments, with only few exceptions (e.g.,
Kroupa 2002).  This robust relation is therefore
recognized as a fundamental diagnostic of the massive star formation
process.

As a follow-on to Ant's model for the log-normal region of the IMF
(A. Whitworth, these Proceedings), it turns out that
the Salpeter slope can be explained as a simple result
of a self-similar, hierarchical star-formation process, based on
successive generations of fragmentation into an $M^{-2}$ mass
distribution.  The mass distribution of clusters and OB associations
is observed to follow this mass distribution (e.g., Elmegreen \&
Efremov 1997; Zhang \& Fall 1999), as well as the {\sc Hii} region 
luminosity function, which best reflects the zero-age cluster mass
function (e.g., Oey \& Clarke 1998).  Even sparse associations and
groups of high-mass stars show this smooth $\alpha = -2$ power law
down to individual O stars in the Small Magellanic Cloud (Oey et
al. 2004).  Furthermore, the mass function of giant molecular clouds
and star-forming clumps within them are also known to be consistent
with $\alpha = -2$, as seen, for example, in presentations at this
meeting (e.g., S. Pekruhl, and others in these Proceedings).  In
contrast, the Salpeter slope is slightly 
steeper, having a value of $\alpha=-2.35$ instead of --2.  

The $M^{-2}$ power law is a reasonable distribution to
expect for the initial mass function of these hierarchical
quantities.  It is the power-law exponent which describes the mass
equipartition between high and low-mass objects.  Furthermore, as shown by
Zinnecker (1982), a cloud with a random mass distribution
of proto-stellar seeds will produce an $m^{-2}$ stellar IMF if the
seeds simply grow by Bondi-Hoyle accretion as  $\dt{m}\propto m^{2}$, until
the entire cloud is absorbed into the stellar masses.  And, Cartwright
\& Whitworth (2012; and these Proceedings) point out that the IMF
should follow a stable distribution function which results from the
sum of random variables.  They show that the core mass function can be
described by such a function, the Landau distribution, which has a
--2 power-law tail.  

\begin{figure}[t]
\includegraphics[scale=.5]{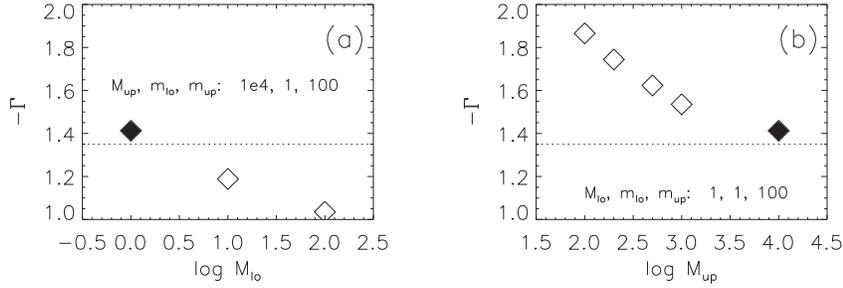}
\caption{Dependence of logarithmic IMF slope $\Gamma$ on the
lower and upper clump mass limits.}
\label{f_montecarlo}       
\end{figure}

It is therefore natural to believe that the hierarchical fragmentation
of molecular clouds into clumps and clumps into stars takes place self
similarly according to a --2 power law mass distribution, therefore
implying that the true, raw stellar IMF has this $\alpha=-2$ relation.
So why is the observed Salpeter IMF slightly steeper?  The answer lies
in the mass range of the stars ($\mlo$ to $\mup$) relative to that of
their parent clumps ($\Mlo$ to $\Mup$).  If a cluster is generated from a
single cloud, then its IMF is that for the aggregate of all stars
formed out of all the clumps in this cloud.  These clumps are
described by a --2 power law.  If $\Mlo < \mup$, then the smallest
clumps are too small to produce the highest-mass stars, thus
slightly suppressing the formation of the highest-mass stars for the
aggregate cluster.  It turns out that the Salpeter slope results for
the condition $\Mlo\sim\mlo$ and $\Mup\gg \mup$ (Oey 2011).

Figure~\ref{f_montecarlo}$a$ shows the results of Monte Carlo
simulations of cluster populations generated by drawing both clump and
stellar masses from a power law with slope $\alpha=-2$.  We assume
a stellar mass range of $\mlo = 1\ \msun$ to $\mup = 100\ \msun$, and
a high upper-mass limit for the clumps, $\Mup = 10^4\ \msun$.
Figure~\ref{f_montecarlo}$a$ shows the dependence of the logarithmic
IMF slope $\Gamma = \alpha + 1$ as a function of lower clump mass
$\Mlo$.  At the highest value of $\Mlo = 100\ \msun$, the mass ranges
for the stars and clumps do not overlap, and essentially all stellar
masses can be formed in all clumps.  We therefore see that the IMF has
the same value as its raw, input slope.  But as $\Mlo$ decreases to values
$< \mup$, the formation of the highest-mass stars is suppressed, since
they can no longer form in the smaller clumps.  This steepens the
aggregate IMF slope.  We see that a value close to the Salpeter slope
(shown by the dotted line) results for the condition $\Mlo = \mlo$.

The models in Figure~\ref{f_montecarlo}$b$ keep $\Mlo = 1\ \msun$
fixed as $\Mup$ decreases.  The black points in
Figures~\ref{f_montecarlo}$a$ and $b$ represent the same model for
which the stellar mass range is 1 -- 100 $\msun$ and the clump mass
range is 1 -- $10^4\ \msun$.  We see that the IMF slope 
continues to steepen, approaching a value near $\Gamma = -1.9$ when
$\Mup = \mup$ and $\Mlo = \mlo$.  Thus, the stellar mass range and
clump mass range are exactly coincident for that model.  Oey (2011)
discusses the effect of additional parameters.

The Monte Carlo simulations show that the Salpeter slope corresponds
to the particular condition that $\Mlo = \mlo$ and $\Mup \gg\mup$.
The condition for $\Mup$ is reasonable, but is $\Mlo \le
\mlo$?  We note that the Salpeter slope applies only to the upper-mass
tail of the IMF, and it is truncated at the lower-mass end by the
observed turnover near 1 $\msun$.  This feature is generally believed
to be linked to the Jeans mass, or in any case, some physics that is
not scale-free.  Therefore, the relevant lower clump mass is that
which produces 1 $\msun$ stars.  And in fact, we do know that
the power-law, clump mass function extends down to $1\ \msun$.
Note that we have discussed this analysis in terms of a 100\% star
formation efficiency.  However, the results are
independent of the star formation efficiency, provided that it is
constant across all clump masses.  Thus, $M$ represents the clump mass
capable of forming that total stellar mass, rather than the physical
clump mass itself.  Therefore, the relevant physical  $\Mlo$ is much larger
than 1 $\msun$ for star formation effiencies $< 100$\%.

\section{Supporting Evidence}

Since we observe the clump mass function to have a power law
distribution to well below the masses needed to form individual 1
$\msun$ stars, this therefore implies that if star formation is indeed
a hierarchical process, then the resulting aggregate IMF for an entire
star cluster {\bf must} be steeper than the raw IMF because of the
suppression of the highest stellar masses in the smallest clumps.
The observed Salpeter slope of $\alpha=-2.35$ cleanly implies such a
steepening from a raw IMF having $\alpha=-2$, which we argued above is
an eminently reasonable value to expect from first principles.

Other observations are also consistent with this model.  As seen in
Figure~\ref{f_montecarlo}, our simulations show that any real scatter in
the IMF should be limited between values of roughly $\Gamma = -1$ to
--2.  This is indeed the
range seen in the observed IMF slopes, as shown by Kroupa (2002).

In addition, starbursts are sometimes suggested to have somewhat
flatter IMF slopes.  The Arches cluster near the Galactic center is
the best-studied example, showing a slope of $\Gamma = -1$ (Espinoza
et al. 2009; Kim et al. 2009).  This flattening can
be understood if starbursts are forming stars so intensely that the
stars form faster than the cloud fragmentation timescale.  Thus the
starburst IMF directly reflects the raw IMF, rather than an aggregate
formed out of cloud fragments.  In other words, we could think of the
entire starbursting cloud as a single giant, star-forming clump.

Finally, this steepening of the aggregate IMF slope relative to
component sub-regions is in fact seen in the large-scale numerical
simulations of Bonnell et al. (2003, 2008).  As shown by Maschberger et
al. (2010), the IMF slope steepens from $\alpha\sim -1.9$ to --2.2
between the subregions and the total aggregrate in the simulation
totaling $10^4\ \msun$, in agreement with our predictions in Figure
~\ref{f_montecarlo}$a$. 

\section{Conclusion}

We stress that a model of hierarchical star formation {\bf must} lead
to steepening of the aggregate IMF slope if the star-forming clumps
have masses $\Mlo \ll \mup$ (Oey 2011).  We know this condition to be true
empirically within star clusters.  If the hierarchical process is 
self-similar, then this implies that the Salpeter slope $\alpha=-2.35$
results from a clump mass function having $\alpha=-2$, which is a
reasonable slope to expect from first principles.  This scenario is
supported by both observations and theoretical simulations.

%


\begin{acknowledgement}
I'm grateful to the conference organizers for the opportunity to
present this work, which was supported by the National Science Foundation, grant
AST-0907758 and visitor support from the Instite of Astronomy,
Cambridge. 

\end{acknowledgement}
%

%
%
%

\end{document}